\documentclass[conference]{IEEEtran}
\IEEEoverridecommandlockouts

\usepackage{adjustbox}
\usepackage{cite}
\usepackage{amsmath,amssymb,amsfonts}
\usepackage{algorithmic}
\usepackage[linesnumbered,ruled]{algorithm2e}
\usepackage{amsthm}
\usepackage{array}
\usepackage{graphicx}
\usepackage{textcomp}
\usepackage{paralist}
\usepackage{multirow}
\usepackage{url}
\usepackage{setspace}
\usepackage{tabu}
\usepackage{tablefootnote}
\usepackage{romannum}
\usepackage{enumitem}
\usepackage{arydshln}
\usepackage[normalem]{ulem}
\usepackage{threeparttable}
\usepackage{url}
\usepackage{xspace}
\usepackage{soul}
\usepackage{booktabs}

\usepackage{bm}  
\usepackage{subcaption}
\usepackage[utf8]{inputenc}
\usepackage[dvipsnames]{xcolor}
\usepackage[colorlinks=true,citecolor=black,linkcolor=black, urlcolor=black]{hyperref}
\usepackage{cleveref}
\usepackage{float}
\usepackage{arydshln}
\usepackage[most]{tcolorbox}

\usepackage[]{collab}

\def\name{{Hades}\xspace}
\def\company{Huawei Cloud}
\def\A{Dataset $\mathcal{A}$\xspace}
\def\B{Dataset $\mathcal{B}$\xspace}
\def\C{Dataset $\mathcal{C}$\xspace}
\def\ind{Dataset $\mathcal{B}$ and $\mathcal{C}$\xspace}

\definecolor{background}{rgb}{0.94, 0.97, 1.0}
\definecolor{edge}{rgb}{0.32, 0.48, 0.72}

\crefname{section}{§}{§§}
\Crefname{section}{§}{§§}
\begin{document}
\title{Heterogeneous Anomaly Detection for Software Systems via Semi-supervised Cross-modal Attention}

\author{
  \IEEEauthorblockN{
    Cheryl Lee\IEEEauthorrefmark{1},
    Tianyi Yang\IEEEauthorrefmark{1},
    Zhuangbin Chen\IEEEauthorrefmark{2},
    Yuxin Su\IEEEauthorrefmark{2}\thanks{Yuxin Su is the corresponding author.},
    Yongqiang Yang\IEEEauthorrefmark{3}, and
    Michael R. Lyu\IEEEauthorrefmark{1}
  }

  \IEEEauthorblockA{\IEEEauthorrefmark{1}Department of Computer Science and Engineering, The Chinese University of Hong Kong, Hong Kong, China.\\
    Email: cheryllee@link.cuhk.edu.hk, \{tyyang, lyu, zbchen\}@cse.cuhk.edu.hk}

  \IEEEauthorblockA{\IEEEauthorrefmark{2}Sun Yat-sen University, Guangzhou, China.
    Email: suyx35@mail.sysu.edu.cn}

  \IEEEauthorblockA{\IEEEauthorrefmark{3}Computing and Networking Innovation Lab, Cloud BU, Huawei.
    Email: yangyongqiang@huawei.com}
}


\maketitle
\begin{abstract}
Prompt and accurate detection of system anomalies is essential to ensure the reliability of software systems.
Unlike manual efforts that exploit all available run-time information, existing approaches usually leverage only a single type of monitoring data (often logs or metrics) or fail to make effective use of the joint information among different types of data. Consequently, many false predictions occur.
To better understand the manifestations of system anomalies, we conduct a systematical study on a large amount of heterogeneous data, i.e., logs and metrics.
Our study demonstrates that logs and metrics can manifest system anomalies collaboratively and complementarily, and neither of them only is sufficient.
Thus, integrating heterogeneous data can help recover the complete picture of a system's health status.
In this context, we propose \name, the first end-to-end semi-supervised approach to effectively identify system anomalies based on heterogeneous data.
Our approach employs a hierarchical architecture to learn a global representation of the system status by fusing log semantics and metric patterns.
It captures discriminative features and meaningful interactions from heterogeneous data via a cross-modal attention module, trained in a semi-supervised manner.
We evaluate \name extensively on large-scale simulated data and datasets from \company.
The experimental results present the effectiveness of our model in detecting system anomalies.
We also release the code and the annotated dataset for replication and future research.
\end{abstract}


\section{Introduction}
Recent years have witnessed the scale and complexity of software systems expand dramatically. However, anomalies are inevitable in large-scale software systems, resulting in considerable revenue and reputation loss \cite{lossReport}.
The core competence of service providers stems from guaranteeing the reliability of software systems, where automated anomaly detection is a primary step and has been picked up extensively within the community.
In real-world software systems, many types of monitoring data, including metrics, logs, alerts, and traces, play an essential role in software reliability engineering~\cite{zbIMS, IMSexp}.
In particular, metrics and logs have been widely used for anomaly detection.
Metrics (e.g., response time, number of threads, CPU usage) are real-valued time series measuring the system status.
Logs are semi-structured text messages printed by logging statements to record the system's run-time status.

Tremendous efforts have been devoted to detecting anomalies automatically since manual troubleshooting is impractical and error-prone.
Some approaches rely on metrics~\cite{Adsketch, SRCNN, OmniAnomaly}, while others rely on logs~\cite{LogRobust, PLElog, Deeplog}.
However, as a single source of information is often insufficient to depict the status of a software system precisely, existing methods could produce many false predictions~\cite{huaweidata}
The popularity of large-scale distributed systems worsened the situation, where anomaly patterns are more complex. 
Intuitively, combining different sources of monitoring data can allow fuller utilization of run-time information to analyze the system status holistically.

To verify our intuition, we study the characteristics of system anomalies incurred by typical faults based on a large amount of heterogeneous monitoring data. The data are generated via fault injection on Apache Spark, where we run various workloads and inject 21 typical types of faults. 
Compared to existing open-access datasets, e.g.,~\cite{Challenge2021, huaweidata, fault2, Loghub}, ours possesses temporally aligned heterogeneous run-time information with rich semantics and annotations (i.e., abnormal or not).

Based on the study, we obtain three interesting findings:
\begin{itemize}[topsep=0pt, leftmargin=12pt]
\item Though the presence of critical logs often indicates problems, their absence does not necessarily imply a healthy system status. An important reason is that sometimes determining where and how to place an informative log statement is difficult~\cite{logging}.
\item In some cases, faults do not affect metrics, while in other cases, metrics exhibit unusual patterns (e.g., jitters) even if the system is experiencing minor performance fluctuations instead of faults. Hence, simply identifying anomalous metric patterns is insufficient.
\item Faults can cause unexpected behaviors involving either logs or metrics, or both of them. So the two data sources should be analyzed comprehensively to reveal the actual anomalies.
\end{itemize}
These findings necessitate considering heterogeneous data for anomaly detection, and high-level patterns (i.e., log semantics and metric patterns) of the observed data deserve full attention.

However, we identify three challenges in extracting and integrating essential information from heterogeneous data.
\textit{(1)~Complex intra-modal information.} 
Logs reflect system anomalies mainly through their semantics (e.g., keywords) and sequential dependencies across events.
Besides, metrics reflect diverse aspects of the system status, and metrics of different aspects tend to develop distinct behavior patterns. For example, when a system works normally, the disk usage often moves steadily, while the CPU usage can fluctuate dramatically. 
Such complex and diverse data patterns call for a model to be highly competent in information processing and feature extraction.
\textit{(2)~Significant inter-modal gap.} 
Logs and metrics are in different forms, i.e., textual and time series. Such a discrepancy poses a huge challenge to effectively using the joint information for downstream anomaly detection. To this end, it is critical to align the log semantics and metric patterns.
\textit{(3)~Trade-off between cost and accuracy.}
Supervised approaches are effective but require high-quality labels.
Annotating massive logs and metrics is prohibitively difficult, costly, and time-consuming, so the requirement of annotations is usually the bottleneck of putting supervised approaches into practice.
Though unsupervised learning avoids labeling by mining inherent trends of unlabeled data to discover anomalies, it suffers inaccuracies with less human oversight and ignorance of valuable domain expertise.

To tackle these challenges, we propose \textbf{\name}, a \underline{H}eterogeneous \underline{A}nomaly \underline{DE}tector via \underline{S}emi-supervised learning for large-scale software systems, equipped with a novel cross-modal attention mechanism.
The key idea is to learn a discriminative representation of the system status based on logs and metrics with limited labeled data for training.
\name first captures intra-modal dependencies using a hierarchical architecture. Then it generates a global representation of the most discriminative latent information of logs and metrics via a modal-wise attentive fusion module.

Specifically, \name involves four components for data modeling and engages semi-supervised training. The components are:
(1) For logs, we adopt the FastText algorithm~\cite{fasttext} and Transformer~\cite{Attention} to model lexical semantics and sequential dependencies of logs.
(2) We employ a hierarchical encoder to learn metric representations based on the causal convolution network~\cite{TCN}. It jointly learns aspect-oriented temporal dependencies, cross-metric relationships, and inter-aspect correlations.
(3) We design a novel modal-wise attention mechanism to facilitate learning meaningful intra- and inter-modal properties.
(4) Finally, the framework infers the system status and triggers an alarm upon detecting anomalies.
The semi-supervised training comprises two phases: 
First, we apply a few labeled data to train the initial model and then pseudo-label the remaining unlabeled data via the current training model. 
Second, the model is updated using both labeled and pseudo-labeled data with high confidence until convergence.

We evaluate \name using one simulated and two datasets from \company.
The experimental results demonstrate the superiority of \name, which achieves an average F1-score of 0.933 and outperforms all state-of-the-art competitors, including log-based and metric-based ones.
Extensive ablation experiments further confirm the effectiveness of our designs (i.e., exploiting heterogeneous information, cross-modal learning, attentive fusion, and intra-modal feature extraction).

In summary, the main contributions of this paper are:

\begin{itemize}[topsep=0pt, leftmargin=12pt]
\item We systematically study how heterogeneous data manifest system anomalies. To our best knowledge, we are the first to point out the collaborative and complementary relationship of logs and metrics in manifesting anomalies.

\item We propose the first end-to-end semi-supervised approach, \name, to effectively detect system anomalies based on heterogeneous monitoring data via cross-modal attention. 

\item Extensive experiments on simulated data and datasets from \company ~demonstrate the effectiveness of \name, as well as the contribution of each design to \name.

\item We collect an annotated dataset containing complex log semantics and metric patterns, which is released with our code for this study~\cite{Hades} of this work to facilitate other related practitioners and researchers.
\end{itemize}
\section{Problem Statement}\label{sec:background}
\begin{figure}[htb]
    \centering
        {\includegraphics[width=0.95\linewidth]{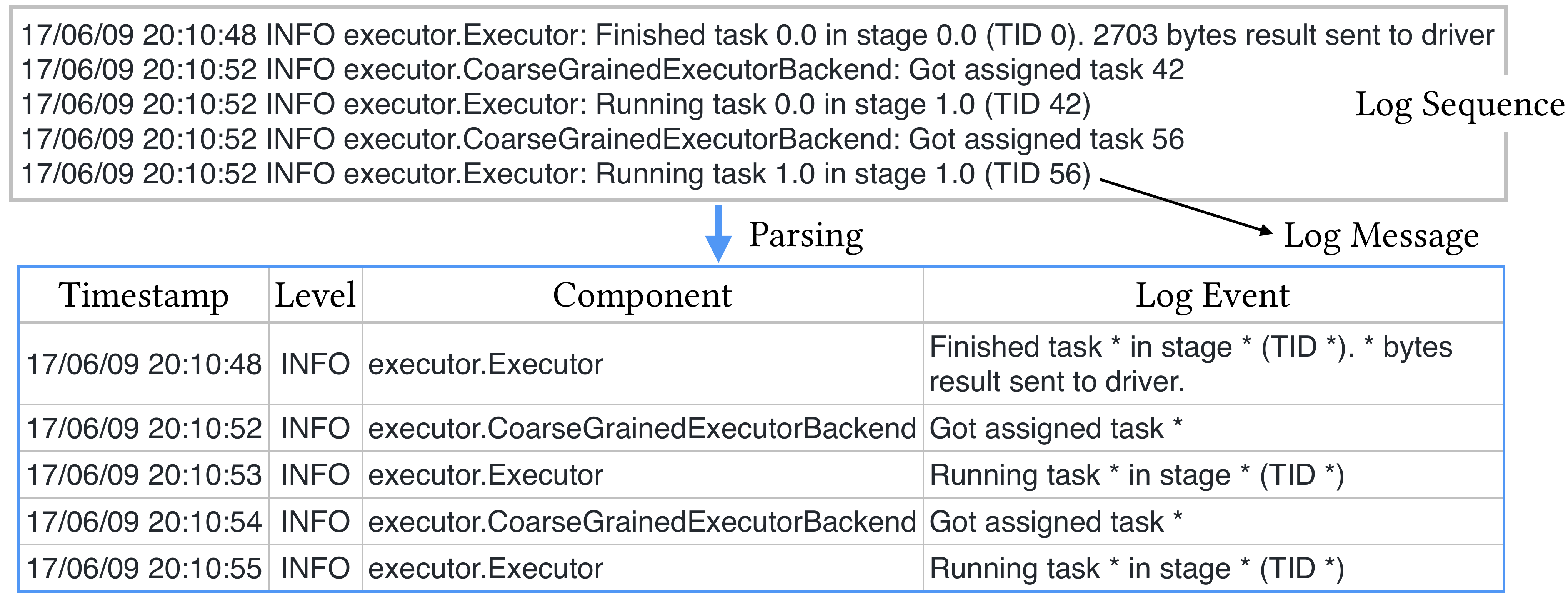}}
    \caption{Examples of logs and parsed logs.}
    \label{fig:log}
 \vspace{-0.1in}
\end{figure}

We first introduce essential terminologies.
A \textit{log message} is a line of the standard output of logging statements, composed of constant strings (written by the developers) and variable values (determined by the system)~\cite{POP}.
Parsing a log message is to remove all variables to obtain a \textit{log event}, which describes system run-time events~\cite{SurveyHe}.
Log messages chronologically collected within a certain period constitute a \textit{log sequence}.
Figure~\ref{fig:log} shows an example of logs and obtained log events.
On the other hand, \textit{metrics} are the numerical measurement of system performance that are sampled uniformly. Consecutive points within a certain period make up a \textit{metric segment}.
By collecting both the logs and metrics in a given period of length $T$, we obtain a \textit{chuck} with time-aligned heterogeneous data. The value of $T$ is determined according to real-world requirements.
\textit{Anomalies} are abnormal system behaviors, events, or observations that do not conform to the expected patterns in the run-time information~\cite{SurveyHe}. These anomalies often indicate system issues and could evolve into errors or failures.

We formalize the heterogeneous anomaly detection task and present this learning problem with inputs and outputs.
Given a data chunk, we aim to determine the current system status as abnormal or normal, denoted as 1 and 0, respectively.
Let $\langle \boldsymbol{X}_{1:N}, Y_{1:N} \rangle=\{(\boldsymbol{X}_1, y_1), (\boldsymbol{X}_2, y_2), \cdots, (\boldsymbol{X}_N, y_N)\}$ be the training data chunks with corresponding status labels, where $\boldsymbol{X}_i=(\boldsymbol{X^l}_i, \boldsymbol{X^m}_i), (i=1,2,...,N)$ is the $i$-th data chunk, and $y_i \in \{0, 1\}$ denotes the status.
In the $i$-th chunk, $\boldsymbol{X^l}_{i,1:L}=[\boldsymbol{x^l}_{i,1}, \boldsymbol{x^l}_{i,2}, ..., \boldsymbol{x^l}_{i,L}]$ denotes the log sequence, where $L$ is the number of log events generated during the period of length $T$;
the metric segment is denoted by $\boldsymbol{X^m}_{i,1:T}=[\boldsymbol{x^m}_{i,1}, \boldsymbol{x^m}_{i,2},$ $..., \boldsymbol{x^m}_{i,T}] \in \mathbb{R}^{T \times M}$, where $M$ and $T$ are the number and length of monitoring metrics, respectively.
The goal is to model the relations behind $\langle \boldsymbol{X}_{1:N}, Y_{1:N} \rangle$, and then for each incoming unseen instance ${X}_{N+1}$, we can predict the status $y_{N+1}$.
\section{Data Collection}\label{sec:data}
In practice, engineers usually analyze different sources of system run-time information for troubleshooting.
However, manual inspection is tedious and fallible, especially when facing massive data. To explore the opportunity to automate this process, we systematically study how system anomalies affect the different types of monitoring data.
Moreover, most industrial datasets are access-restricted, and the publicly accessible data are often too small or single-source due to security and privacy concerns. To alleviate this problem, we have released our collected dataset on~\cite{Hades}.
It contains large-scale logs and metrics generated from a distributed computing system, which underpins our study and will facilitate the advancement and openness of the community.

Our data collection comprises four steps: 1) deploy the infrastructure, 2) conduct workloads to generate monitoring data, 3) inject typical faults to simulate industrial production anomalies, and 4) collect heterogeneous monitoring data simultaneously. We describe them in detail as follows.

\subsection{Data Generation}
Apache Spark is a widely-used framework for big data processing~\cite{Spark}. We deploy Spark 3.0.3 on a distributed systems cluster containing a master node and five worker nodes (virtual nodes supported by Docker~\cite{Docker}) in our laboratory environment.
To conduct workloads, we employ a big data benchmark suite, HiBench~\cite{HiBench}. 
Unlike existing collections (e.g.,~\cite{Loghub, fault1, fault3}) running only word counting, we perform 19 kinds of workloads covering diverse service application scenarios.
The workloads fall into five categories and are diverse in terms of resource usage, including CPU, I/O, memory, and network.
For example, the workload random forest performs complex computation, and the workload word counting requires transaction-intensive file inputting.
Details about the settings of data collection are introduced in~\ref{sec:appendix:data}.

We first repeat each workload without faults seven times, where the run-time data are named as \textit{standard} data when no fault is injected during the entire workload. 
We mainly collect two types of run-time data, i.e., logs and metrics. Logs are aggregated by Spark automatically, and metrics are sampled (per second) and collected via an open-source monitoring tool~\cite{PAT}.
In the following analysis, we focus on 11 monitoring metrics reflecting four critical aspects of the system status (i.e., CPU, I/O, memory, and network), such as CPU system usage, device read speed (``rkb/s''), memory usage, and network throughput rate.

\subsection{Fault Injection}\label{sec:study:fault}
After gathering standard data, we inject 21 typical types of faults that are selected based on previous research~\cite{fault1, fault2, fault3} and our investigation of the typical service failures at \company. 
They fall into four categories:
\begin{itemize}[leftmargin=12pt, topsep=0pt]
\item Process suspension: suspend a process for one minute, including the Application Worker, Master, Node Manager, Datanode, Resource Manager, and Namenode.
\item Process killing: kill each process once, including the Application Worker, Master, Node Manager, and Datanode. Killing these processes does not terminate the running workload immediately.
\item Resource stress: load the system by injecting CPU, memory, and other resource hogs, supported by~\cite{stress-ng}.
\item Network faults: inject Linux traffic faults such as losing packets, high latency, and flashing disconnections.
\end{itemize}
Each fault lasts for one minute, and the time from fault injection to fault clearance is called \textit{fault duration}. The data collected for the rest of the time are regarded as \textit{fault-free}.

\subsection{Data Annotation}\label{sec:study:annotation}
We invite two Ph.D. students experienced in software reliability as annotators. They are unaware of when the faults are injected or cleaned, so the labeling is completely based on the data without off-site information.
The principle to label an anomaly is that the in-process data manifest discrepancies from the standard data, e.g., error logs and unusual metric jitters.
Meanwhile, the abnormal manifestations should align with the expected impact of the injected fault, which is manually checked by annotators.
We do not simply regard all data generated during the fault duration as abnormal because many faults can be immediately tolerated by the system's fault-tolerant mechanisms, incurring no anomalies to the system. In this case, we treat the corresponding data as normal.
The label is accepted if the two students give the same label for one chunk independently. Otherwise, they will discuss with a post-doc until reaching a consensus. The inter-annotator agreement~\cite{cohen1960coefficient} (i.e., a measure of the reliability of an annotation process) achieves 0.876 before adjudication.
\section{Motivation}\label{sec:study}
This section elaborates on our motivation by answering the following three questions:
\begin{itemize}[leftmargin=*, topsep=0pt]
    \item How do logs manifest system anomalies?
    \item How do metrics manifest system anomalies?
    \item How do monitoring data (including logs and metrics) reflect the system status?
\end{itemize}

\subsection{How do logs manifest system anomalies?}\label{sec:study:log}
Logs may not be susceptible enough to some system faults. Only 3.62\% of positively labeled chunks are anomalous from the log's perspective.
A typical example provides a closer look. If the network drops some packets, the service response becomes slow but may not hit the pre-defined timeout threshold, and thereby no anomalous log event will be reported. 
The main reason lies in the inherent deficiency of logs.
Logging~\cite{he2018characterizing, yuan2012improving, yuan2012conservative, SurveyHe} is a human activity heavily relying on developers' knowledge. While it is relatively easy to write logs describing severe failures, subtle performance issues, e.g., gray failures~\cite{huang2017gray}, are often hard to identify. Thus, logs capturing such anomalies could be missing.

Digging into logs, we observe that the lexical semantics are noteworthy.
Specifically, the appearance of some log tokens indicates anomalies.
These tokens only occur in abnormal sequences, and their semantics describe unusual system events, e.g., ``uncaught'' and ``exception'' in the event ``Uncaught exception in thread''.
This observation is in line with the programming habits of developers, as well as previous studies~\cite{deep-loglizer, LogRobust, SemParser}. It also validates that log semantics can reflect the current system status to some extent.

Moreover, the contexts of logs are crucial for detecting anomalies since logs carry the information of program control flows~\cite{workflow}. 
For example, a normal sequence of log events is an event reporting the ``final application'' state followed by the log event ``Shutdown hook called''.
When anomalies happen, the event ``Shutdown hook called'' may occur before reporting the ``final application'' state.
In this case, the application state will be regarded as undefined or failed because the master has not received the message informing the application state.
Hence, the contextual semantics of logs also contain information that indicates whether the system is healthy.

\begin{tcolorbox}[colback=background!50,
        colframe=edge,
        width=\columnwidth,
        boxrule = 0.3mm,
        top = 3pt, bottom=3pt, left=3pt, right=3pt
    ]
    \textbf{Finding 1}: Logs sometimes cannot record fine-grained information and therefore, cannot manifest all system anomalies. Moreover, both lexical and contextual semantics are important with respect to reflecting the system status.
\end{tcolorbox}

\subsection{How do metrics manifest system anomalies?}\label{sec:study:metrics}
Metrics are responsive to anomalies by continuously recording system status. However, there are still some anomalous periods ignored by many existing metric-based detectors.
\begin{figure}[htb]
    \centering
    \vspace{-0.1in}
        {\includegraphics[width=0.9\linewidth]{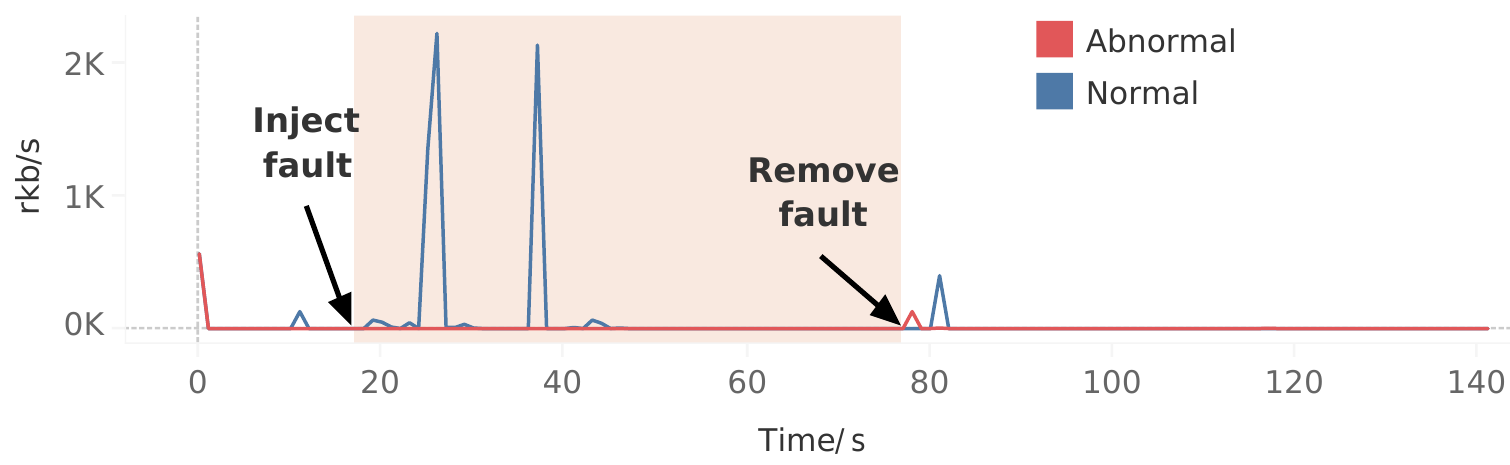}}
    \vspace{-0.1in}
    \caption{Suspending the Datanode incurs anomalies manifested in the metric ``rkb/s'', but no novel pattern exists.}
    \label{fig:in_fault_normal}
\vspace{-0.1in}
\end{figure}

Existing anomaly detectors usually try to identify novel metric patterns through a comparison with normal system behaviors, i.e., novelty detection~\cite{novelty}. They mainly focus on local patterns (e.g., spikes, level shifts) rather than global patterns spanning the entire workload. However, system faults are not necessarily manifested by local novel patterns. 
Figure~\ref{fig:in_fault_normal} displays an example that the metric ``rkb/s'' during the fault ``Datanode suspension'', as shown by the red line.
The metric ``rkb/s'' is abnormal as a whole, but its local patterns are hard to identify as anomalies.
Specifically, the red line remains zero after fault injection, which is unexpected as there should have been I/O activities going on. The blue line depicts the normal (standard) status as a comparison.
However, the blue line also remains zero after the 90-\textit{th} second because the system does not exchange data near the end of the workload. 
So ``rkb/s'' in either the normal or abnormal status can stay at zero for a while.
This indicates that a metric can behave very similarly when the system is in the opposite status.
Such patterns (which may reflect the opposite system statuses) will confuse most existing methods relying on novel pattern mining, leading to performance degradation.
The inherent reason is that system metrics cannot completely reflect the software's inner execution logic.
Fortunately, such anomalies can be detected by referring to the logs, where suspicious events like \textit{Exception in createBlockOutputStream} are reported.

\begin{figure}[htb]
    \centering
    \vspace{-0.1in}
        {\includegraphics[width=0.9\linewidth]{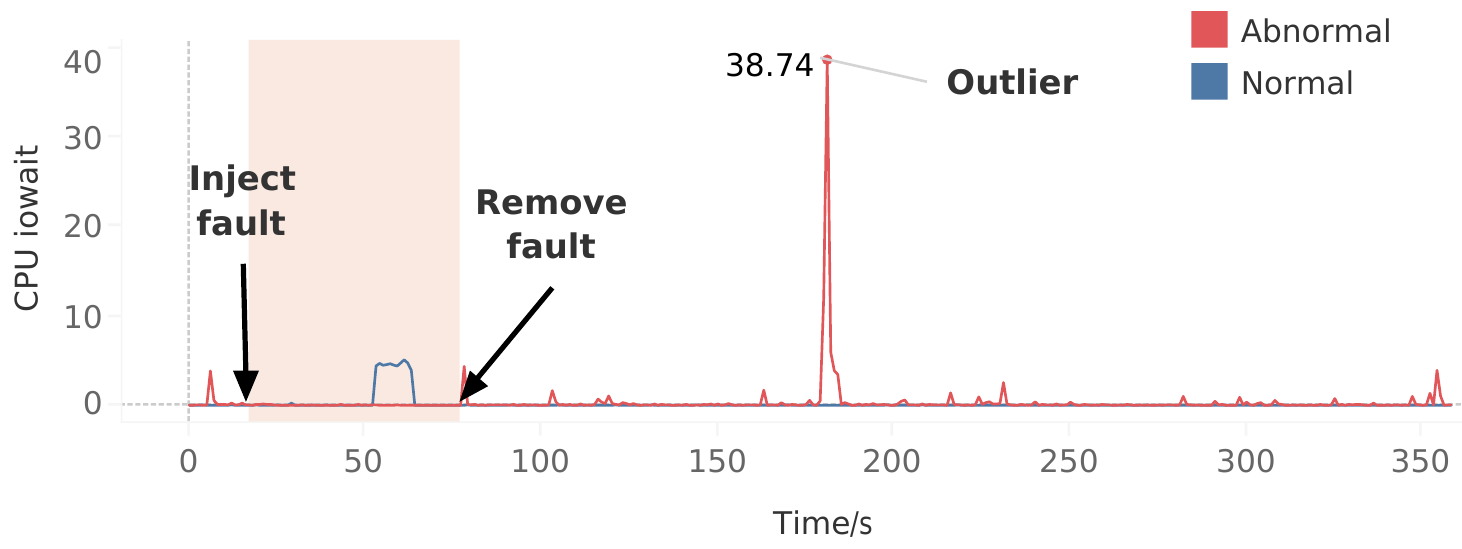}}
    \caption{An acceptable outlier of ``CPU iowait'' in a fault-free period may trigger a false alarm by most automated tools.}
    \label{fig:out_fault_abnormal}
\vspace{-0.1in}
\end{figure}

On the flip side, unusual metric fluctuations may trigger alarms even when no anomaly exists currently. 
Among all metric segments manually labeled as positive, 8.87\% of them are collected in fault-free periods, indicating that relying solely on metrics may cause false alarms due to the overreactions of metrics.
Figure~\ref{fig:out_fault_abnormal} displays an example that the metric ``CPU iowait'' generates a rare heartbeat spike even in the fault-free period.
Such sporadic and transient fluctuation is acceptable without affecting the service, thereby no alarm should be triggered.
This case suggests that other information should be involved to mitigate the issues caused by the over-sensitivity of metrics to avoid unnecessary engineering resource waste. 
\begin{table*}[htb]
\small\centering
\caption{Typical faults and the corresponding anomalous manifestations of logs and metrics}
\vspace{-0.1in}
 \begin{tabular}{|c|c|c|}
\hline
\textbf{Faults} & \textbf{Anomalies in logs} & \textbf{Anomalies in metrics}\\
\hline
Memory hog & Warnings (reaches the memory limit) & Memory-related metrics rise steeply \\
\hline
Virtual memory hog & Errors (reporter thread fails) & CPU and memory-related metrics jitter \\
\hline
I/O hog & Warnings (slow ReadProcessor) & I/O-related metrics rise steeply \\
\hline
Network delay & Warnings (executor heartbeat timeout) & Network-related metrics suddenly drop \\
\hline
Connection flash & Nothing \textbf{(silent)} & Network-related metrics suddenly drop and quickly restore\\
\hline
Datanode killed & Errors (excluding datanode) & Related metrics plummet to zero \textbf{(silent)}\\ 
\hline
Secondary namenode killed & Errors (failed to connect to $<$IP$>$) & Related metrics plummet to zero \textbf{(silent)}\\ 
\hline
    \end{tabular}
    \label{tab:abnormal_behavior}
\end{table*}

\begin{figure}[htb]
    \centering
    \vspace{-0.1in}
        {\includegraphics[width=0.9\linewidth]{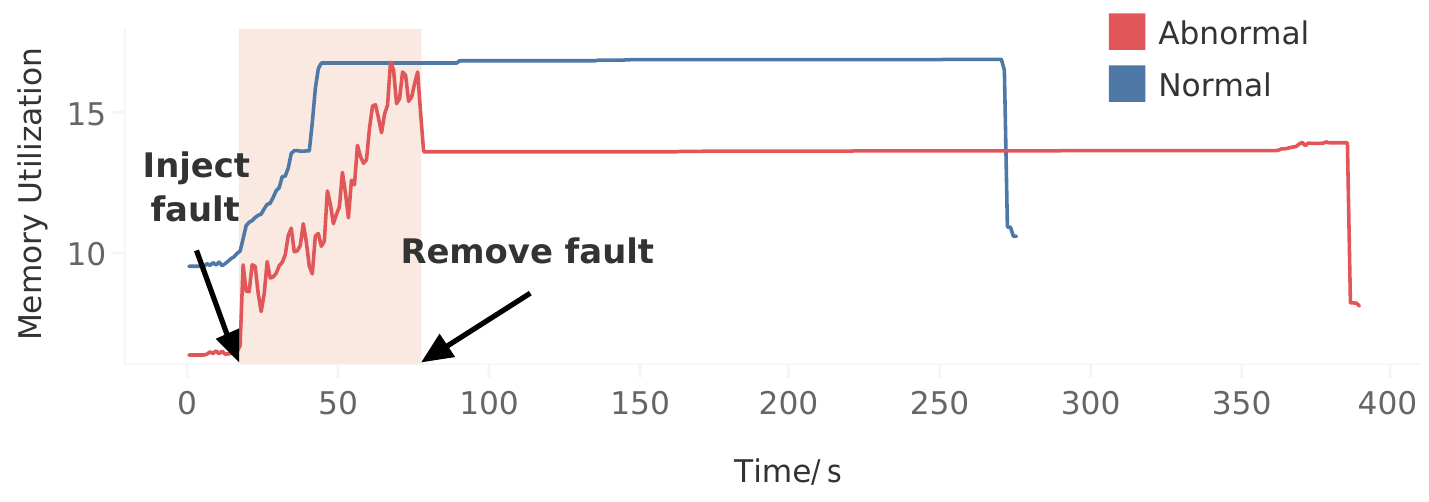}}
    \caption{The irregular metric jitters cannot be detected by single point-based detectors.}
    \label{fig:vm jitter}
\vspace{-0.2in}
\end{figure}

In addition, we find that metric patterns presented at the segment level (i.e., a series of continuous metric points) sketch issues much better than single-point outliers.
For example, the metric ``memory usage'' performs noticeably abnormal jitters when injecting a ``virtual memory hog'' (Figure~\ref{fig:vm jitter}).
However, these jittering points are not outliers because their values are not extraordinarily high or low. Clearly, the outlier perspective ignores high-order data variations, such as the scope and denseness, resulting in missing alarm issues.

\begin{tcolorbox}[colback=background!50,
        colframe=edge,
        width=\columnwidth,
        boxrule = 0.3mm,
        top = 3pt, bottom=3pt, left=3pt, right=3pt
    ]
    \textbf{Finding 2}: Metrics are responsive to system status changes but still insufficient in many cases. Their over-sensitivity may cause false alarms on uncommon yet acceptable fluctuations. Besides, segment-level metric patterns can be more useful in anomaly detection than single-point outliers.
\end{tcolorbox}

\subsection{How do monitoring data reflect the system status?}\label{sec:study:fuse}
We summarize representative faults and their effects on logs and metrics in Table~\ref{tab:abnormal_behavior}, showing that faults can affect logs and metrics simultaneously.
For example, one resource hog can incur both sudden spikes in metrics and warning logs for limited resources.
Moreover, some cases see an evident complementary relationship between logs and metrics.
When the Datanode or Secondary Namenode is killed, related metrics plummet to zero but cannot be detected (i.e., silent) since these metrics also plummet to zero if the application ends normally. Metric-based anomaly detectors cannot distinguish such abnormal drops from normal ones. In this case, logs play the role of additional information to help determine the system status.
On the contrary, when the connection between nodes flashes, no warnings or errors are generated in logs since the network disconnection time is too short to affect program operation.
Nevertheless, network-related metrics faithfully reflect such transient anomalies, e.g., the network throughout that drops rapidly during flashes.
Such observations align with intuition. Basically, logs record the software's internal execution logic, while metrics provide an external view by measuring software services' performances, resource usage, etc.
Thus, combining the two can better portray the system status due to their collaborative and complementary relationships.

Furthermore, a fault can affect logs and metrics to varying degrees.
For example, killing the Datanode during the Latent Dirichlet Allocation (LDA) application causes 29 abnormal metric segments while only one log sequence reports anomalies.
Another example is that when suspending the Namenode in word counting, the related metrics experience a sharp drop and remain unchanged since most of the computation has been done. Yet tens of logs reporting a failed state are generated because the worker nodes keep sending warnings while the master node cannot receive messages.
Hence, simply regarding all types of data equally important is unreasonable since the more severely affected part may deserve more attention.
In brief, we should combine and assign appropriate weights to metrics and logs to promote effective anomaly detection.

\begin{tcolorbox}[colback=background!50,
        colframe=edge,
        width=\columnwidth,
        boxrule = 0.3mm,
        top = 3pt, bottom=3pt, left=3pt, right=3pt
    ]
    \textbf{Finding 3}: Metrics and logs can both respond to anomalies, but neither is sufficiently informative. They have collaborative and complementary relationships in providing clues for the system's health. Also, the degree to which they are affected by the same anomaly can vary greatly.
\end{tcolorbox}

These findings support our motivation to develop an automated anomaly detector based on heterogeneous monitoring data, i.e., logs and metrics. 
This results in HADES, our solution to attack the previously mentioned challenges.
\section{Methodology}\label{sec:hades}
\begin{figure*}[htb]
    \centering
        {\includegraphics[width=0.95\linewidth]{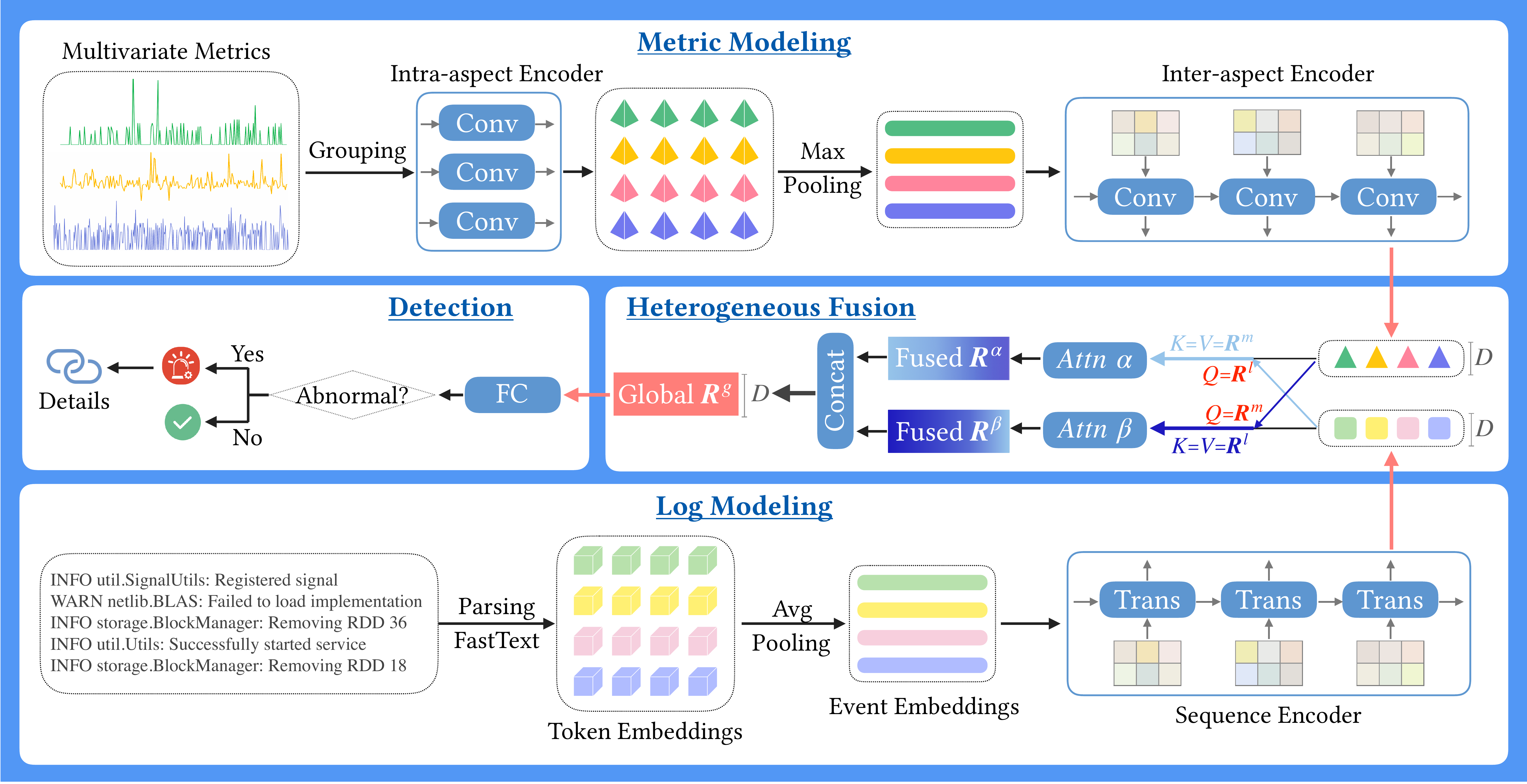}}
    \vspace{-0.08in}
    \caption{Overview of \name.}
    \label{fig:hades}
 \vspace{-0.1in}
\end{figure*}
Figure~\ref{fig:hades} presents the overview of \name, a heterogeneous anomaly detector for software systems via cross-modal attentive learning. It consists of four components: Log Modeling, Metric Modeling, Heterogeneous Representation Fusion, and Detection. It is trained by both labeled and pseudo-labeled data in a Semi-supervised Training manner.
\name aims to infer the system status from current heterogeneous monitoring data based on historically extracted patterns. We incorporate domain knowledge and the insights obtained from our previous study to design a practical model architecture.
Specifically, for logs, \name captures lexical and contextual log semantics and maps each raw log sequence into a low-dimensional representation.
For metrics, \name preserves aspect-aware information at the segment level along the timeline and learns cross-aspect correlations. 
Our framework also employs an attention-based fusion module with cross-modal learning to acquire a global representation, which is fed into a successive detection component. Consequently, it will trigger an alarm to operations engineers when a noteworthy anomaly occurs.

\subsection{Log Modeling}\label{sec:method:log}
Log modeling contains three steps: log parsing, log vectorization, and log representation learning. It extracts meaningful log features including lexical and contextual semantics.

\subsubsection{Log Parsing}
In this step, we transform unstructured log messages into structured log events.
As aforementioned, raw log messages are unstructured and contain variables that can hinder log analysis~\cite{POP}. 
Therefore, we first employ a widely-used parser Drain~\cite{Drain} to extract log events since it has shown effectiveness and efficiency in the previous evaluation study~\cite{parserBenchmark}. 
Next, we conduct a stable sorting based on the log timestamps. 
As a result, all valid log messages are transferred into chronologically arranged log events.

\subsubsection{Log Vectorization}\label{sec:method:log:vector}
This phase turns textual log events into semantics-aware numerical vectors.
We utilize FastText~\cite{fasttext} to capture the intrinsic relationships of log vocabulary and preserve the important log semantics.
FastText is a popular, lightweight, and efficient technique for producing word embeddings that can represent semantic similarities between words. 
After training, FastText maps every token into a $E$-dimension vector, so a log event $\boldsymbol{x^l}$ is transformed into a token embedding list $\boldsymbol{V}= \{v_i\}_{i=1}^{\omega}\in \mathbb{R}^{\omega \times E}$, where $\omega$ is the token number of an event.
Subsequently, we average all elements inside $\boldsymbol{V}$ to acquire a sentence embedding $\bar{\boldsymbol{V}}=\frac{1}{\omega}\times \sum_{i=1}^{\omega} \boldsymbol{v}_i$. 
Consequently, a log sequence $\boldsymbol{X^l}_{1:L}$ can be denoted by a sentence embedding list $\bar{\boldsymbol{V}}=\{\bar{\boldsymbol{V}}_i\}_{i=1}^{L} \in \mathbb{R}^{L \times E}$.

\subsubsection{Log Representation Learning}
This step models the log contextual semantics and generates log representations with learned information.
In particular, the sentence embeddings of a sequence obtained from the previous phase are fed into a sequence encoder, which is composed of two Transformer encoder layers~\cite{Attention}. This encoder captures contextual dependencies across the events.
Afterward, a fully-connected (FC) layer maps the output into a $D$-dimensional feature space. 
Hence, we obtain the log representation of a chunk, denoted by $\boldsymbol{R^{l}} \in \mathbb{R}^{L \times D}$. 
If the log sequence is too long, we partition it into non-overlapping fixed-size sub-sequences and conduct the above steps.
For a too-short sequence, we pad it with zeros.

\subsection{Metric Modeling}\label{sec:method:metric}
Metrics are modeled hierarchically in this section with respect to segment-level patterns motivated by the second finding in~\cref{sec:study:metrics}. The module comprises an intra-aspect encoder and an inter-aspect encoder (also shown in Figure~\ref{fig:hades}).

The rationale behind the design is three-fold:
(1) Metrics sketching the same aspect of the system should be modeled together.
Monitoring metrics can reflect various aspects of system performance, e.g., CPU utilization, memory utilization, etc.
Generally, metric patterns of the same aspect share certain similarities (e.g., CPU user usage and CPU system usage both characterize CPU usage).
Such metrics should be grouped and regarded as multi-variate time series (MTS) to be analyzed together.
(2) Metrics depicting different aspects should be modeled separately.
If two metrics belong to different aspects, their patterns can be very different. For example, the disk usage tends to be stable while the I/O throughput may fluctuate violently even under the normal status.
So metrics of different aspects should be fed into separate models to capture fine-grained information.
(3) While metrics of different aspects tend to develop distinct patterns, they still exhibit some inter-aspect correlations when anomalies occur.
For instance, if a worker node loses connection with the master node, many metrics such as the CPU utilization and route cost will drop precipitously and stay at zero since data exchange or computation is interrupted.

Based on (1) and (2), we propose an intra-aspect encoder to capture the aspect-oriented temporal dependencies and cross-metric relationships.
(3) inspire us to design an inter-aspect encoder to learn correlations across aspects integrally.
The two encoders learn the aspect-aware representations hierarchically.

\subsubsection{Intra-Aspect Encoder}
As metrics should be modeled in an aspect-aware manner, that is, modeling metrics of the same aspect together while modeling metrics of different aspects separately, the internal model must be computationally efficient.
Thus, we adopt 1D causal convolution~\cite{TCN}. Conventional convolution networks face the problem of information leakage (i.e., the output depends on future inputs) and the inability of sequential dependency modeling. Causal convolution is designed to mitigate these limitations, which meets our needs by being parallelizable, lightweight, and accurate~\cite{TCNexp}.
This phase decomposes the metrics into $\gamma$ groups according to their corresponding aspects based on our domain knowledge.
After that, metrics of the same aspect are taken as an MTS and fed into a separate intra-aspect encoder composed of a multi-layer causal convolution network. 
After appropriate padding and chomping, the $\gamma$ intra-aspect encoders output $\gamma$ feature vectors $\boldsymbol{h^{m}}$.
Finally, we conduct a max-pooling operation on the feature dimension and stack the outputs to form a latent feature vector $\boldsymbol{H^{m}} \in \mathbb{R} ^{T \times \gamma}$.

\subsubsection{Inter-Aspect Encoder}
This module also leverages causal convolution to learn the inter-aspect features. Such structure helps model complex patterns by capturing multi-level information.
We take $\boldsymbol{H^{m}}$ outputted by the intra-aspect encoder as an MTS with ${T \times \gamma}$ series and feed it into the inter-aspect encoder to model the correlations between metric aspects.
In this way, the metrics $\boldsymbol{X^{m}}_{1:T}$ inside a chunk are embedded into a $D$-dimensional representation, denoted by $\boldsymbol{R^{m}} \in \mathbb{R}^{T \times D}$.
Note that data of all modalities should be embedded into the same $D$-dimensional feature space for alignment.

\subsection{Heterogeneous Representation Fusion}\label{sec:method:fusion}
We design a novel cross-modal attention mechanism to fuse heterogeneous representations and bridge the gap between logs and metrics.
Previous phases embed logs and metrics into a $D$-dimensional feature space. These representations are fed together into this fusion module, defined by two attention layers~\cite{Attention}.
The first one (Attn-$\alpha$) takes the log representation $\boldsymbol{R^{l}}$ as the Query while the metric representation $\boldsymbol{R^{m}}$ as the Key and Value. It matches the log events explaining the metric changes.
Symmetrically, in the second attention layer (Attn-$\beta$), $\boldsymbol{R^{m}}$ plays the role of the Query, and $\boldsymbol{R^{l}}$ serves as the Key and Value. It helps to find the performance variations aligned with log contents to enhance log expressiveness.
Mathematically, given the Query, Key, and Value, we calculate:
\begin{equation}\label{eq:attn}
\mathsf{Fuse}(Q, K, V) = \mathsf{tanh} \left( \left[\mathsf{softmax} (QW_sK^{\mathrm{T}})V; Q\right] W_a \right)
\end{equation}
where $W_a$ and $W_s$ are learnable parameters; $[\cdot; \cdot]$ denotes concatenation.
Afterward, outputs from Attn-$\alpha$ and Attn-$\beta$ are concatenated inside the $D$-dimensional space to constitute a global representation $\boldsymbol{R^{g}} \in \mathbb{R}^{(T+L)\times D}$, defined as:
\begin{equation}
\label{eq:global}
\boldsymbol{R^{g}} = [\mathsf{Fuse}(\boldsymbol{R^{l}}, \boldsymbol{R^{m}}, \boldsymbol{R^{m}}); \mathsf{Fuse}(\boldsymbol{R^{m}}, \boldsymbol{R^{l}}, \boldsymbol{R^{l}})]
\end{equation}
Above all, switching the roles of the two modalities allows devoting more attention to the features of different modalities that convey similar information, which is more likely to be responsive to changes in the system status.
Also, it can retain meaningful intra-modal patterns explicitly by directly concatenating the Query with the attended Value. In this way, the global representation reserves not only the shared information and cross-modal interactions but also the salient intra-modal dependencies and the inferred features due to the complementary relationship between logs and metrics.

\subsection{Detection}\label{sec:method:detection}
\begin{figure}[htb]
    \centering
    \vspace{-0.1in}
        {\includegraphics[width=0.99\linewidth]{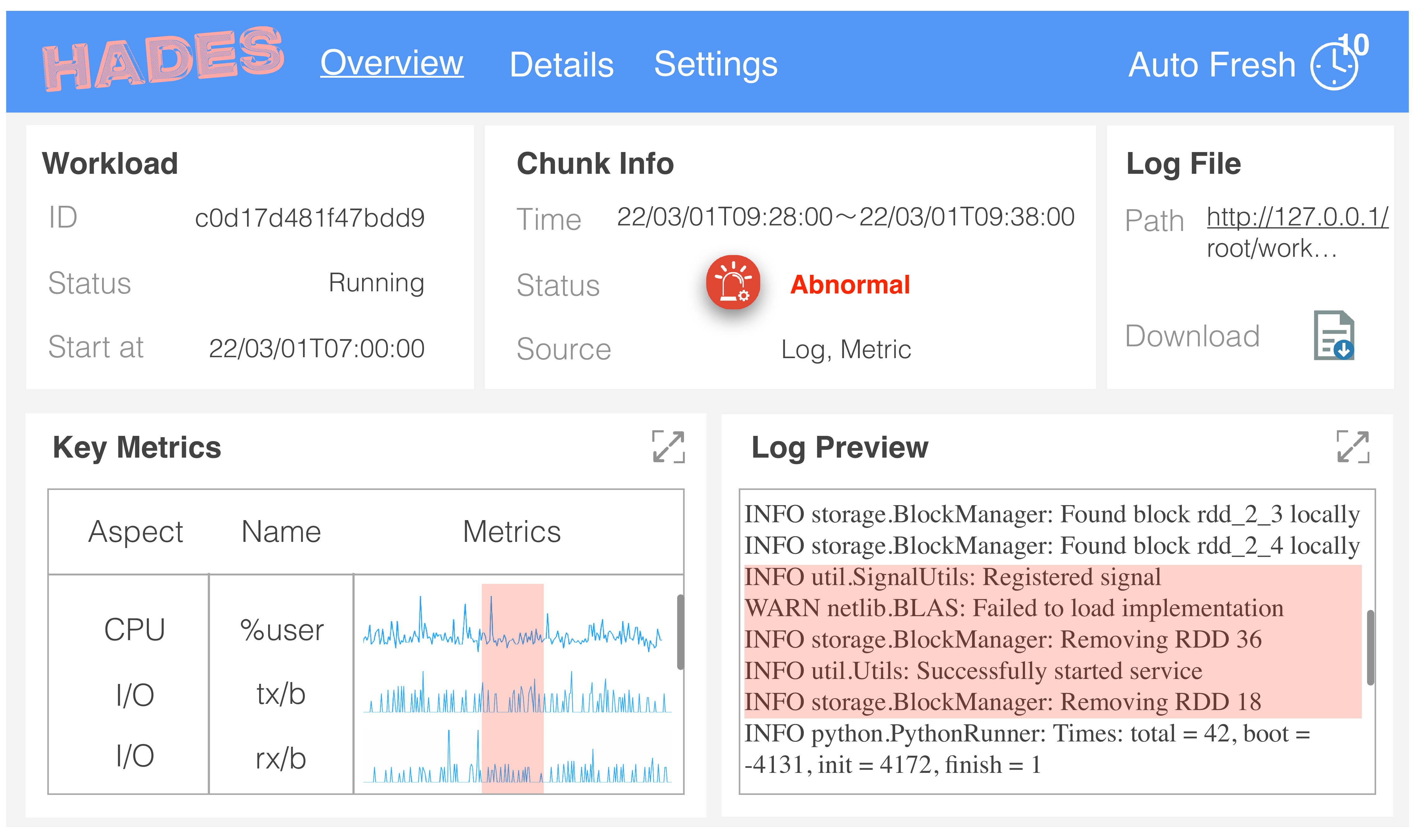}}
    \vspace{-0.1in}
    \caption{A demo for reviewing the detected chunk, where the light red rectangle represents the currently focused window.}
    \label{fig:interface}
\vspace{-0.1in}
\end{figure}

Finally, we feed the representation $\boldsymbol{R^{g}}$ into stacked FC layers followed by a softmax layer. The output $\hat{y} \in \{0,1\}$ represents the status being normal or abnormal, computed by:
\begin{equation}
\label{eq:res}
\hat{y} = \mathsf{argmax}[\mathsf{softmax}(U\sigma(V\cdot \boldsymbol{R^g}+b)+c)]
\end{equation}
where $U$ and $V$ are learnable weight matrices; $b$ and $c$ are bias terms; $\sigma(\cdot)$ is the ReLU activation function~\cite{ReLU}.
This module generates an alarm upon detecting an anomaly.
To facilitate further analysis by engineers, we also provide a visual interface that enables a convenient review of logs and metrics of the suspicious chunk, as shown in Figure~\ref{fig:interface}.

\subsection{Semi-supervised Training}\label{sec:method:training}
\begin{figure}[htb]
    \centering
    \vspace{-0.1in}
        {\includegraphics[width=0.8\linewidth]{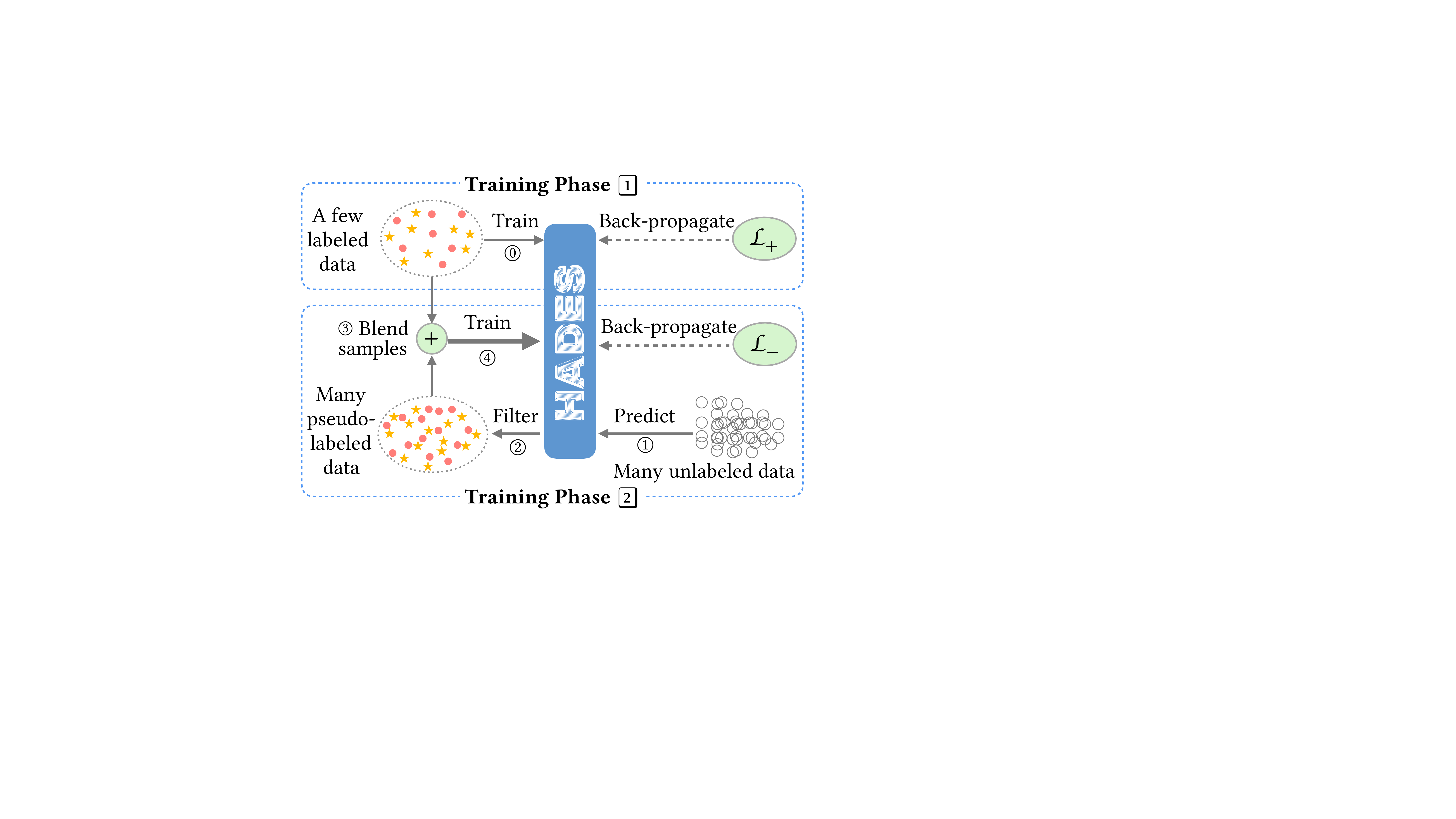}}
    \vspace{-0.1in}
    \caption{The semi-supervised training of \name.}
    \label{fig:semi}
\vspace{-0.1in}
\end{figure}

To reduce the cost of manual labeling and leverage human expertise simultaneously, we apply semi-supervised learning to train our model. Semi-supervised learning leverages a small amount of labeled data and unlabeled data for training, based on the smoothness assumption: 
a normal sample should be closer to another normal sample rather than an abnormal sample in the feature space, and vice versa.
Previous studies adopt the assumptions that logs with similar semantics should share the same detection indicator~\cite{PLElog}, and metrics with similar patterns lead to the same existence indicator of performance issues~\cite{Adsketch}. Intuitively, heterogeneous data with similar log semantics, metric patterns, and cross-modal interactions are likely to share the same label. With this intuition, we devise a two-fold semi-supervised training strategy, shown in Figure~\ref{fig:semi}.

We first label data generated during frequent faults by representative workloads (accounting for about 10\% of the whole data). The labeled data are used to train \name in a supervised manner in the first phase by minimizing the binary cross-entropy loss (denoted as $\mathcal{L}_{+}$) for several epochs.
Next, we generate predictions of the remaining unlabeled data and filter in predictions with high confidence as pseudo labels. As the model is a probabilistic classifier in nature and thereby provides confidence scores directly, we take the maximum of the outputted classification probabilities as the corresponding prediction's confidence.
The second phase blends pseudo-labeled and labeled data to train the framework, attempting to minimize the following loss function:$\mathcal{L} = (1-\lambda) \cdot \mathcal{L}_{+} + \lambda \mathcal{L}_{-}$, where $\mathcal{L}_{-}$ is the binary cross-entropy loss of the pseudo-labeled data, and $\lambda$ is a hyper-parameter ranging $[0,1]$.
\section{Evaluation}\label{sec:experiment}
We evaluate \name by answering the following research questions (RQs):
\begin{itemize}[leftmargin=*, topsep=0pt]
\item \textbf{RQ1}: How effective is \name in anomaly detection?
\item \textbf{RQ2}: What is the contribution of each design of \name?
\item \textbf{RQ3}: How sensitive is \name to the length of a chunk?
\end{itemize}

\subsection{Experiment Setup}
\subsubsection{Datasets}
Besides the in-lab \A (\cref{sec:data}), we also evaluate \name on other two datasets ($\mathcal{B}$ and $\mathcal{C}$) containing heterogeneous monitoring data from two different industrial cloud services of \company.
\ind contain eight types of faults respectively: CPU stress, memory stress, high disk I/O latency, disk partition full, network flashing, long network latency, high packet loss rate, and zombie process. 
Metrics are sampled once per minute. Drain~\cite{Drain} extracts 72 and 104 log events from \ind, respectively.

Chunks are obtained via a timestamp-based sliding partitioning strategy~\cite{deep-loglizer}. For all three datasets, we split the first 70\% generated data as the training set and the last 30\% as the test set, meaning that data in different sets are produced in different periods (also different workloads for \A).
Inside the training set, we choose five representative workloads out of 19 in $\mathcal{A}$ (choose one in each category according to~\cite{HiBench}) and seven faults out of 21 for labeled training data, that is, only annotations of the chosen workloads during the standard status or the selected faults' duration are adapted for training. The rest of the training set is taken as unlabeled training data.
For \ind, we invite three experienced reliability engineers to annotate the test set, and training data with two selected faults (memory stress and disk partition full), following a similar procedure described in~\cref{sec:study:annotation}. Few disagreements occurred as anomalies are well-defined by the engineers who are familiar with their system.

In this way, chunks belonging to the same faulty or non-faulty period will exist either in the training/test set so as to avoid possible data leakage caused by random split. Besides, the test set contains unseen anomalies incurred by non-selected faults, so we can evaluate the model's ability to infer new abnormal patterns, considering knowing all the anomaly patterns in advance is usually impossible.
Table \ref{tab:datasets} shows the statistics of our datasets. 
\begin{table}[htb]
\small
\centering
\caption{Dataset statistics}
\vspace{-0.1in}
\begin{adjustbox}{max width=\columnwidth}
\begin{tabular}{cccc}
\toprule
Dataset & Log Messages & Metric Length & Manually Labeling  \\
\midrule
$\mathcal{A}$ & 1,435,139 & 64,422 & 10.87\% \\
$\mathcal{B}$ & 76,724 & 7,473 & 13.70\% \\ 
$\mathcal{C}$ & 1,148,563 & 15,945 & 11.24\% \\ 
\bottomrule
\end{tabular}\label{tab:datasets}
\end{adjustbox}
\vspace{-0.1in}
\end{table}

\subsubsection{Baselines}
We compare \name with ten baselines in different training manners, including unsupervised, semi-supervised, and supervised ones. 
In particular, despite being specially designed for bad software identification, SCWarn~\cite{SCWarn} (unsupervised) is the only multi-source approach for binary classification in the software engineering literature, as far as we know.
Six state-of-the-art single-source baselines are adopted: Deeplog~\cite{Deeplog} (unsupervised), PLELog~\cite{PLElog} (semi-supervised), and LogRobust~\cite{LogRobust} (supervised) are log-based. Omninomaly~\cite{OmniAnomaly} (unsupervised), Adsketch~\cite{Adsketch} (semi-supervised), SRCNN (unsupervised) and its supervised variant SRCNN-s~\cite{SRCNN} are metric-based.
We also adopt supervised log-based SVM (SVM-$\mathcal{L}$) and metric-based SVM (SVM-$\mathcal{M}$)~\cite{SVM} as the representation of traditional techniques.

\subsubsection{Evaluation Measurements}
As we tackle anomaly detection in a binary classification manner, we adopt the widely-used measurements to gauge models' performances: 
\textit{Rec}$=\frac{TP}{TP+FN}$, \textit{Pre}$=\frac{TP}{TP+FP}$, \textit{F1}$=\frac{2\cdot TP}{2\cdot TP+FN+FP}$, 
where \textit{TP} is the number of successfully detected abnormal chunks; 
\textit{FP} is the number of normal chunks incorrectly triggering alarms;
\textit{FN} is the number of undetected abnormal chunks.

\subsubsection{Implementations}
Our code for implementing \name is publicly available at~\cite{Hades}.
The log encoder adopts a hidden size of 1024 for four layers. We use Gensim~\cite{Gensim} to train 32-dimensional word embeddings.
The intra-aspect metric encoder comprises two layers, and the inter-aspect encoder has three layers with a hidden size of 256.
The decoder consists of four layers with a hidden size of 512.
We use the Adam optimizer~\cite{Adam} with an initial learning rate of 0.001.
The batch size is 128, and the epoch for each training phase is 50.

As for baselines, we adopt the public implementations of~\cite{SCWarn, OmniAnomaly, Adsketch, PLElog, SRCNN}, and an open-source toolkit~\cite{deep-loglizer} to implement~\cite{LogRobust, Deeplog}, whose original paper did not provide code.
We determine the hyper-parameter combination achieving the highest test \textit{F1} for each baseline.
All experiments are conducted on a single NVIDIA GeForce GTX 1080 GPU.

\subsection{RQ1: Overall Performance of \name}
\begin{table*}[htb]
\small
\centering
\vspace{-0.1in}
\caption{Overall Performance Comparison.}
\vspace{-0.1in}
\begin{tabular}{*{10}{c}}
\toprule
\multirow{2}*{\textbf{Models}} & \multicolumn{3}{c}{\A} & \multicolumn{3}{c}{\B} & \multicolumn{3}{c}{\C} \\
\cmidrule(lr){2-4}\cmidrule(lr){5-7}\cmidrule(lr){8-10}
&\textit{F1} & \textit{Rec} & \textit{Pre}
&\textit{F1} & \textit{Rec} & \textit{Pre}
&\textit{F1} & \textit{Rec} & \textit{Pre}\\
\cmidrule{1-10}\morecmidrules\cmidrule{1-10}
SCWarn & 0.321 & 0.389 & 0.273 & 0.497 & 0.643 & 0.405 & 0.491 & 0.585 & 0.423\\
\hdashline[0.5pt/5pt]
SVM-$\mathcal{L}$ & 0.289 & 0.707 & 0.181 & 0.541 & 0.756 & 0.421 & 0.481 & 0.742 & 0.356\\
DeepLog & 0.259 & 0.386	& 0.194 & 0.386 & 0.526 & 0.305 & 0.375 & 0.524 & 0.292\\
PLELog & 0.314 & 0.602 & 0.213 & 0.463 & 0.618 & 0.371 & 0.434 & 0.597 & 0.341\\
LogRobust & 0.404 & 0.684 & 0.287 & 0.524 & 0.718 & 0.413 & 0.495 & 0.698 & 0.383\\
\hdashline[0.5pt/5pt]
SVM-$\mathcal{M}$ & 0.536 & 0.833 & 0.395 & 0.608 & 0.839 & 0.477 & 0.556 & 0.801 & 0.426\\
OmniAnomaly & 0.681 & 0.788 & 0.601 & 0.827 & 0.863 & 0.794 & 0.812 & 0.896 & 0.743\\
Adsketch & 0.404 & 0.476 & 0.351 & 0.543 & 0.644 & 0.470 & 0.538 & 0.649 & 0.459\\
SRCNN & 0.342 & 0.614 &	0.237 & 0.467 & 0.701 & 0.350 & 0.472 & 0.586 & 0.394\\
SRCNN-s & 0.784 & 0.826 & 0.745 & 0.898 & 0.938 & 0.861 & 0.883 & 0.926 & 0.844\\
\midrule
\textbf{\name} & \textbf{0.864} & \textbf{0.870} & \textbf{0.859} & \textbf{0.975} & \textbf{0.984} & \textbf{0.966} & \textbf{0.960} & \textbf{0.969} & \textbf{0.951}\\
\bottomrule
\end{tabular}
\vspace{-0.1in}
\label{tab:main experiments}
\end{table*}

Table~\ref{tab:main experiments} presents the overall performance comparison. Each result of \name is averaged over three independent runs with three random seeds and the values of evaluation measurements vary in very small decimal cases.
\name outperforms all baselines by a significant margin, achieving the best result on every evaluation measurement.
Specifically, its \textit{F1} scores are 0.864, 0.975, and 0.960, 9.12\%$\sim$174.41\% higher than competitors on average.
The high \textit{Rec} and \textit{Pre} scores of \name illustrate that there are very few missed anomalies or false alarms.
Thus, we can argue that \name is considerably effective to detect system anomalies, redounding to economizing engineering resources.

Compared with SCWarn, the success of \name boils down to three aspects:
(1) \name adopts semi-supervised learning to balance annotation cost and human oversight. SCWarn is unsupervised and detects anomalies based on the next timestamp prediction, yet accurately predicting is really difficult in a large-scale system with complex behavior patterns. Sometimes software behaviors are inherently unpredictable~\cite{software_behavior}.
(2) \name extracts multi-level log semantics and represents log events via succinct and low-dimensional embeddings.
In comparison, SCWarn transforms logs into event occurrence sequences, generating an over-large sparse feature matrix for log events and discarding log semantics. The sparse matrix poses barriers to extracting meaningful features, especially when hundreds of events exist.
(3) \name devises an attentive fusion to capture significant cross-modal interactions and bridge temporal and textual representations. SCWarn simply concatenates the representations and ignores the vast gap between metrics and logs, i.e., the information form and the input size discrepancies.

The superiority of \name over single-modal baselines stems from the effective use of logs and metrics.
Compared with the best single-modal model (SRCNN-s, a supervised approach), \name increases \textit{F1}, \textit{Rec} and \textit{Pre} by 9.12\%, 4.91\%, 13.22\% on average, respectively.
Such improvement is exciting and reasonable. Our previous study reveals that metrics and logs can both reflect anomalies, and neither of them is sufficient (\cref{sec:study:fuse}).
However, these baselines only review one data source and omit important information hidden in the other source, so they suffer performance degradation, especially for those without knowledge of historical anomalies.

In addition, training an epoch takes \name two minutes on average while baselines take 0.37$\sim$9.01 minutes per epoch. 
The prediction delay is negligible as the prediction of all mentioned approaches is  less than 0.1s for a chunk. The efficiency of \name with nearly real-time detection and acceptable offline training time engages its industrial prospect.

In brief, \name effectively detects system anomalies on all datasets. It significantly improves the effectiveness compared to all baselines concerning every evaluation measurement.

\subsection{RQ2: Individual Contribution of Modules}
\subsubsection{Derived Models}
We conduct an extensive ablation study on \name. Particularly, we derive seven models based on the original \name to investigate the contribution of the introduction of heterogeneous information, representation fusion, cross-modal attention, and intra-modal feature extraction.

\begin{itemize}[leftmargin=*, topsep=0pt]

\item Heterogeneous Information:
\name\textit{w/o}$\mathcal{M}$ removes metrics, containing a log encoder (duplicated from~\cref{sec:method:log}) and a decoder. 
Similarly, \name\textit{w/o}$\mathcal{L}$ removes logs, containing a metric modeling module (duplicated from~\cref{sec:method:metric}) and a decoder.
That is, representations of logs and metrics are directly fed into the respective decoder separately.

\item Representation fusion: \name\textit{w/o}$\mathcal{F}$ performs a Boolean \texttt{OR} operation on the results from \name\textit{w/o}$\mathcal{M}$ and \name\textit{w/o}$\mathcal{L}$. It is built on the motivation that it is natural to process each type of data individually and then aggregate results instead of fusing representations as \name does.

\item Cross-modal attention:
\name\textit{w/o}$\mathcal{A}$ removes the cross-modal attention and simply concatenates the representations of metrics and logs as the global representation. 
\name\textit{w/o}$\mathcal{C}$ operates conventional self-attention on the two representations separately and then concatenates them, rather than using cross-modal attention. 
Other modules except for the fusion module (\cref{sec:method:fusion}) are the same as \name.

\item Intra-modal feature extraction:
\name\textit{w/o}$\mathcal{H}$ is designed for validating the contribution of the hierarchical aspect-aware metric encoder (\cref{sec:method:metric}), which models metrics in an aspect-agnostic manner based on causal convolutions by encoding all metrics simultaneously. 
\name\textit{w/o}$\mathcal{S}$ aims to present the usefulness of log lexical semantics by replacing the word embedding with one-hot encoding. 

\item Semi-supervised training: \name-Anno are trained by annotated data rather than leveraging the semi-supervised training approach, which is used to evaluate the effectiveness gap between our proposed semi-supervised training and supervised training.
\end{itemize}

\begin{table*}[htbp]
\small
\centering
\caption{Experimental Results of the Ablation Study.}
\vspace{-0.1in}
\begin{tabular}{*{10}{c}}
\toprule
\multirow{2}*{\textbf{Models}} & \multicolumn{3}{c}{\A} & \multicolumn{3}{c}{\B} & \multicolumn{3}{c}{\C} \\
\cmidrule(lr){2-4}\cmidrule(lr){5-7}\cmidrule(lr){8-10}
&\textit{F1} & \textit{Rec} & \textit{Pre}
&\textit{F1} & \textit{Rec} & \textit{Pre} 
&\textit{F1} & \textit{Rec} & \textit{Pre} \\
\cmidrule{1-10}\morecmidrules\cmidrule{1-10}
\name\textit{w/o}$\mathcal{M}$ & 0.296 & 0.737 & 0.185 & 0.468 & 0.660 & 0.363 & 0.418 & 0.597 & 0.318\\
\name\textit{w/o}$\mathcal{L}$ & 0.719 & 0.718 & 0.720 & 0.840 & 0.886 & 0.799 & 0.832 & 0.888 & 0.782\\
\hdashline[0.5pt/5pt]
\name\textit{w/o}$\mathcal{F}$ & 0.817 & 0.761 & 0.881 & 0.910 & 0.931 & 0.890 & 0.898 & 0.886 & 0.911\\
\hdashline[0.5pt/5pt]
\name\textit{w/o}$\mathcal{A}$ & 0.829 & 0.831 & 0.828 & 0.943 & 0.966 & 0.921 & 0.928 & 0.931 & 0.926\\
\name\textit{w/o}$\mathcal{C}$ & 0.841 & 0.829 & 0.853 & 0.953 & 0.947 & 0.959 & 0.938 & 0.934 & 0.943 \\
\hdashline[0.5pt/5pt]
\name\textit{w/o}$\mathcal{H}$ & 0.852 & \textbf{0.882} & 0.824 & 0.967 & 0.980 & 0.955 & 0.952 & \textbf{0.970} & 0.935\\
\name\textit{w/o}$\mathcal{S}$ & 0.830 & 0.814 & 0.847 & 0.938 & 0.916 & 0.962 & 0.927 & 0.909 & 0.945\\
\hdashline[0.5pt/5pt]
\name-Anno & \textbf{0.866} & 0.878 & 0.855 & \textbf{0.979} & 0.972 & \textbf{0.986} & \textbf{0.961} & 0.953 & \textbf{0.970} \\
\midrule
\textbf{\name} & 0.864 & 0.870 & \textbf{0.859} & 0.975 & \textbf{0.984} & 0.966 & 0.960 & 0.969 & 0.951\\
\bottomrule
\end{tabular}
\vspace{-0.2in}
\label{tab:ablation}
\end{table*}

\subsubsection{Experimental Results}
Table~\ref{tab:ablation} shows the experimental results, underpinning four key conclusions:
(1) Introducing heterogeneous information contributes incredibly to enhancing anomaly detection in view of two observations.
\name outperforms \name\textit{w/o}$\mathcal{M}$ and \name\textit{w/o}$\mathcal{L}$ considerably, especially in \textit{Pre}, indicating that \name can reduce a large number of false alarms, thereby increasing \textit{F1} by 136.80\% and 17.06\% on average, respectively.
Also, all heterogeneous data-based derived models outperform homogeneous variants (\name\textit{w/o}$\mathcal{M}$ and \name\textit{w/o}$\mathcal{L}$), further confirming that heterogeneous data goes far toward fully characterizing the system health.
(2) Two shreds of evidence highlight the benefits of representation fusion:
1) \name performs better than \name\textit{w/o}$\mathcal{F}$ (6.63\% higher in \textit{F1} on average); 
2) the variants using representation fusion (\name\textit{w/o}$\mathcal{A}$ and \name\textit{w/o}$\mathcal{C}$) outperform \name\textit{w/o}$\mathcal{F}$.
It is not surprising since \name\textit{w/o}$\mathcal{F}$ cannot fully mine cross-modal interactions. For example, the over-sensitivity of metrics (stated in~\cref{sec:study:metrics}) may cause false alarms, and \name\textit{w/o}$\mathcal{F}$ fails to overcome such inaccuracy, while \name alleviates this issue by utilizing logs as supplementary information to make more reasonable inferences.
(3) Our cross-modal attention shows extraordinary value as \name achieves better results than \name\textit{w/o}$\mathcal{A}$ and \name\textit{w/o}$\mathcal{C}$ because \name can filter more informative features and exploit higher-order cross-modal interactions, thereby probing a stronger ability to characterize the system status.
(4) The devised encoders make contributions via fuller- and finer-grained feature extraction, supported by the two observations:
\name\textit{w/o}$\mathcal{H}$ generates more false alarms with mixing patterns of diverse aspects, resulting in overall worse performance (lower \textit{F1});
\name\textit{w/o}$\mathcal{S}$ is relatively ineffective as it ignores crucial lexical semantics of logs.
(5) The applied semi-supervised training is comparatively effective as supervised training while requiring only 10\% labels. \name is only 0.10\%$\sim$0.41\% lower in \textit{F1} than its supervised version.
The success can be attributed to two reasons. 
First, only a small part of labeled data can cover various patterns, encouraging accurate inferences of \name on similar unlabeled data samples.
Second, we adopt a two-phase training strategy so the second semi-supervised training phase only considers pseudo labels with high confidence, thereby avoiding error accumulation during the two phases.

The results not only conform to the findings of our previous study in~\cref{sec:study}, but also reveal the significance of heterogeneous data and the competence of our designs for intra- and inter-modal representation learning.

\subsection{RQ3: Sensitivity to Chunk Length}\label{sec:experiment:sensitivity}
\begin{figure}[hbt]
    \centering
    \vspace{-0.1in}
        {\includegraphics[width=0.96\linewidth]{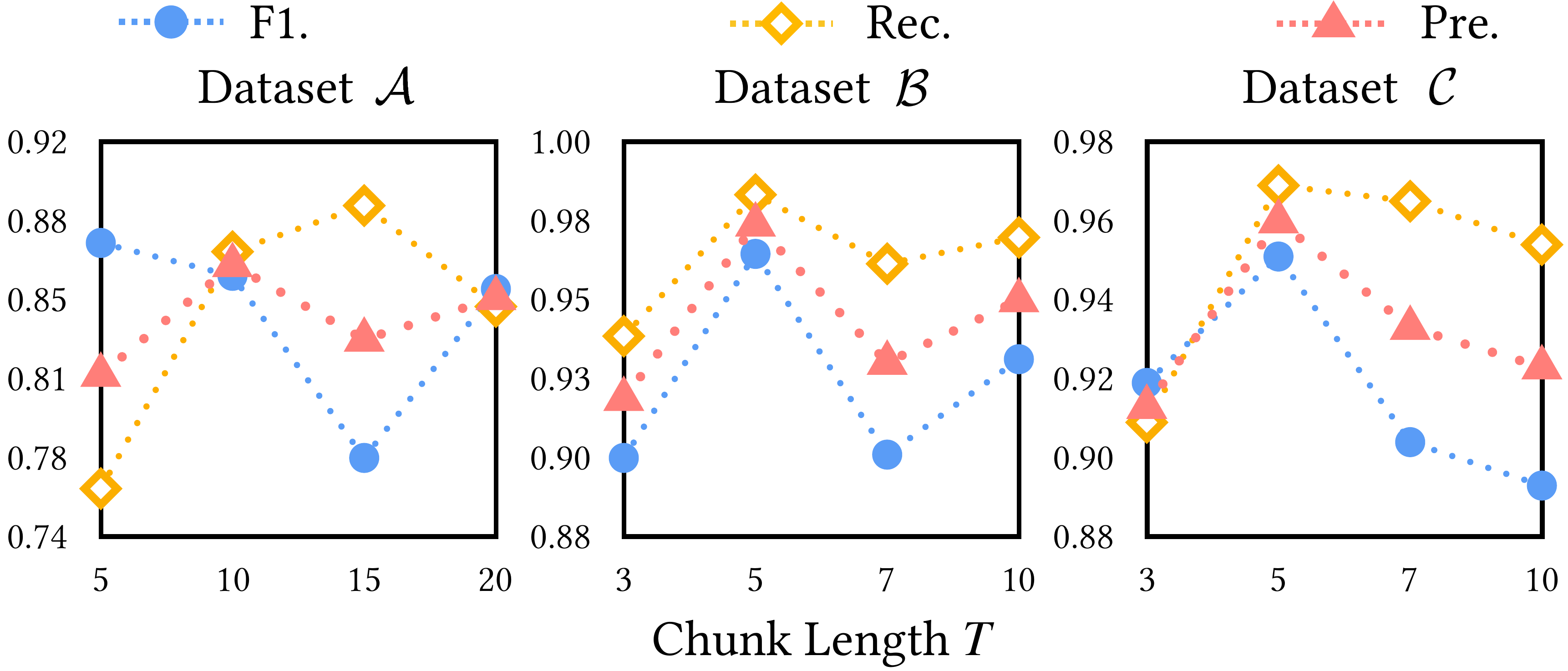}}
    \vspace{-0.05in}
    \caption{The sensitivity to chunk length}
    \label{fig:length}
\vspace{-0.1in}
\end{figure}

The pre-determined chunk length \textit{T} (e.g., the length of a chunk) may affect the framework's performance by affecting the dataset distribution. 
We herein evaluate the sensitivity of~\name to this hyper-parameter.
We change the value of \textit{T} while keeping all other hyper-parameters unchanged and conduct experiments as in RQ4. 
In detail, for \A, the default value of \textit{T} is 10 (sec), ranging from 5 to 20; for \ind, \textit{T} is 5 (min) by default and ranges from 3 to 10, as the sampling frequencies of different datasets are different.
Figure~\ref{fig:length} presents the experimental results.
Overall, \name is fairly stable under different settings of \textit{T}, further confirming its robustness. This makes \name easy to deploy and launch in practice.
We observe that more missing anomalies are rendered when the value of \textit{T} deviates from the default configuration.
This may be because the default granularity of chunks felicitously fits the anomalous patterns. Note that a larger \textit{T} cannot guarantee an easier catch of anomalies. The influence of chunk length depends on multiple factors including model characteristics and system behaviors.
In practice, we do not have to find the ideal \textit{T}. It can be selected according to the validation result or the engineering expertise and requirements.

\section{Discussion}\label{sec:discuss}
\subsection{Lessons Learned}
We argue to be beware of the semantic gap between natural language and logs.
Recent studies~\cite{BertLog, LogRobust, LogAnomaly} have gradually adopted models pre-trained in natural language corpus to encode log texts. However, the semantics of natural language and logs are not exactly identical. 
For example, the word ``successfully'' usually expresses positive emotions in natural languages, yet it indicates an anomaly in the event ``\textit{Successfully connected to}$<$\textit{IP}$>$:$<$\textit{NUM}$>$ for \textit{BP-*-}$<$\textit{*}$>$'', which occurs 
when the Application Worker tries to reconnect with Datanode after losing the connection, and the re-connection succeeds. However, such a re-connection is not expected as connection loss should not happen.
To mitigate this problem, \name uses self-trained word embeddings and integrates multi-level semantics including word-level semantics, event-level semantics, and cross-event sequential dependencies.

\subsection{Threats to Validity}\label{sec:threat}
A potential \textit{internal threat} lies in the acquisition of log event embeddings. We use the average token embeddings as an event embedding, ignoring the sequential information inside a single event, as it is too time-consuming to extract the sequential dependencies event by event.
Nevertheless, based on our study and engineering experience, it is almost impossible for two log events to have identical tokens but in different orders, let alone to represent opposite system statuses.
Thus, omitting the intra-event lexical order will not cause apparent adverse effects.

The \textit{external threat} mainly comes from our datasets. 
Though \name is evaluated on three datasets, it is yet unknown whether the effectiveness of \name can be generalized across other datasets.
To mitigate this threat, we use different datasets with representative workloads and typical faults to evaluate \name, and the experimental results also demonstrate that \name can work well even on unseen anomalies.
We will also evaluate our approach based on more datasets in the future.
In addition, the annotation process may introduce noise. Labeling principles vary depending on the system/person/purpose, sometimes causing inconsistency, especially when it comes to non-extreme anomalies or rare patterns, such as the slightly steep ups and downs in metrics, metric variations between normal fluctuations and abnormal jitters, log statements that rarely occur, etc.
To alleviate this concern, we invite expert annotators that are familiar with the monitoring data. Semi-supervised training is also noise-resistant by only adopting a part of the labels.
In practice, a company will share the definition of anomalies and develop standard labeling rules internally. It will also invite multiple professional practitioners to perform labeling to get labeling results with good consistency.

\subsection{Limitations}\label{sec:limitation}
We identify two limitations.
First, \name is complex and suitable for large-scale systems, such as cloud systems containing plenty of components rather than small systems with simple data patterns. 
Smaller systems may require only naive methods to obtain sufficiently accurate results, so applying \name is not cost-effective.
To improve its efficiency in small systems, we can distill the parameters of \name to form a lightweight version.
Second, \name requires simultaneously collected logs and metrics. However, we find that in some large companies, metrics and logs are collected separately by different departments, and the sampling/logging frequency of different data sources varies dramatically. Sometimes a certain data source is even absent for a while.
Thus, we add an extra mode to allow alternate use of homogeneous or heterogeneous anomaly detection. With either logs or metrics, practitioners can still take advantage of the superior feature extraction capability of \name in the absence of a particular data source, thus extending the applicability of our approach.
\section{Related Work}\label{sec:review}
Many efforts have been devoted to automated anomaly detection for large-scale system reliability insurance based on logs and metrics~\cite{SwissLog, BertLog, Log2Vec, LogAnomaly, EGADS, InterFusion, USAD, TranAD}. 
Many advanced log-based anomaly detectors adopt deep learning to model log sequences.
For example, \cite{Deeplog} and \cite{PLElog} used recurrent neural networks to model normal logs and regard logs deviating from the model as abnormal.
\cite{LogRobust} tackled log instability by introducing the attention mechanism, as well as employing word embedding and TF-IDF.
Metric-based methods usually attempt to model metrics by capturing the temporal dependencies~\cite{Telemanom, EGADS}, mining representative patterns~\cite{Adsketch}, and learning inter-series relationships~\cite{InterFusion, OmniAnomaly}.
\cite{SRCNN} borrowed the spectral residual idea from visual saliency detection, making it easy to use in practice no matter whether labeled data exist. 
Recently, Chen \textit{et al.}\cite{Adsketch} achieved state-of-the-art by discovering anomalous metric patterns that sketch particular performance issues.
Different from \name, these methods only leverage single-source data, ignoring rich information from diverse sources of data and their interactions.  

Some approaches employ multi-source data to identify software rollouts or operations that incur failures~\cite{ADCloud15, Col14, Log3C, Gandalf, SCWarn}.
\cite{ADCloud15, Col14, Log3C} explored the correlations between logs and metrics to identify bad rolling upgrade operations. They regard logs as operation records and metric variations as the consequences of operations, instead of detecting anomalies via neck-and-neck fused information.
\cite{Gandalf, SCWarn} leveraged logs and key performance indicators to alarm bad software changes, whereas \cite{Gandalf} used each source separately rather than analyzing all sources generically, and \cite{SCWarn} transformed logs into event occurrence series, and modeled series together with homogeneous indicators.
However, these identifiers either regard metrics as a posteriori information rather than combing logs and metrics at the same level, or convert heterogeneous data into a homogeneous form ignoring the gap and high-order interactions among different data types. 
\name outperforms these works by learning meaningful cross-modal heterogeneous representations while narrowing the log-metric gap.
\section{Conclusion}\label{sec:conclusion}
We study the manifestations of typical system anomalies on heterogeneous monitoring data and point out for the first time that logs and metrics have collaborative and complementary relationships in reflecting system anomalies, but neither of them only is sufficient.
Motivated by the findings, we propose the first end-to-end semi-supervised approach, \name, to detect anomalies effectively by exploiting heterogeneous information.
\name adopts semi-supervised learning to incorporate human oversight while reducing the annotation cost. It leverages multi-level log semantics and aspect-aware dependencies of metrics, all the while learning meaningful log-metric interactions via cross-modal attention.
We evaluate \name with comprehensive experiments on three datasets, and the results demonstrate that \name outperforms all competitive approaches. Furthermore, the data, code, and experiment results in this study are released for replication.

\section*{Acknowledgement}
The work described in this paper was supported by the National Natural Science Foundation of China (No. 62202511), and the Research Grants Council of the Hong Kong Special Administrative Region, China (No. CUHK 14206921 of the General Research Fund).


\bibliographystyle{IEEEtran}
\bibliography{main}

\appendix
\section{APPENDIX}\label{sec:appendix:data}
\subsection{Workloads}
We conduct 16 workloads on our established cluster based on HiBench~\cite{HiBench}, falling into categories:
\begin{itemize}[topsep=0pt, leftmargin=12pt]
    \item Graph-based: Graph-X-based PageRank and NWeight calculation (an iterative graph-parallel algorithm).
    \item SQL-related operations: Join, Aggregation, and Scan.
    \item Websearch: PageRank.
    \item Basic: Sort (text inputs generated by RandomTextWriter), WordCount, TeraSort, Sleep, and Repartition (benchmarking shuffle performance).
    \item Machine learning: Naive Bayesian Classification, Alternating Least Squares, Latent Dirichlet Allocation, Random Forest, and Singular Value Decomposition 
\end{itemize}
The above workloads cover a large range of general application scenarios to enhance the respectiveness of the collected data.

\subsection{Data}
We collect log files of 37.64MB in total.
After cleaning and parsing these raw log messages with Drain~\cite{Drain}, we obtain 1,048,576 log events with 151 log templates, covering 95.87 hours. The anomaly ratio of log events is 0.055\%. 

The dataset contains 11 metrics with a length of 64,422. Each metric is sampled per second. The metrics include:
CPU usage of User, System, IOwait, and Idle; I/O related metrics including rkB/s, wkB/s, and util; used memory and Commit memory; communication metrics via a network including rxkB/s and txkB/s. The anomaly ratio of metrics is 24.47\%.

\end{document}